\documentstyle[12pt]{article}
\begin{document}
\setcounter{page}{0}
\thispagestyle{empty}

\begin{flushright}
CUPP-95/4; ~  TIFR/TH/95-13  \\
September 1995
\end{flushright}

\begin{center}
{\LARGE\bf SINGLET-HIGGS-BOSON SIGNALS AT HADRON COLLIDERS }
\vskip 5pt
{\it Anindya Datta$^{~1, \heartsuit}$,  Amitava Raychaudhuri$^{~1,
\dagger}$},{\it Sreerup Raychaudhuri$^{~2, \#}$} \\ and \\{\it Surajit
Chakrabarti$^{~3}$}

{\footnotesize $^1$Department of Pure Physics, University of Calcutta,
 \\ 92 Acharya Prafulla Chandra Road, Calcutta 700 009, India.}

{\footnotesize $^2$Theoretical Physics Group,
Tata Institute of Fundamental Research,
\\ Homi Bhabha Road, Bombay 400 005, India.}

{\footnotesize $^3$Department of Physics,
Maharaja Manindra Chandra College,
\\ 20 Ramkanto Bose Street, Calcutta 700 003, India.}

\vskip 5pt
{\large\bf ABSTRACT}
\end{center}
\sl
Many extensions of the Standard Model include $SU(2)_L \times U(1)_Y$
singlet higgs bosons, $h^0$, and also vectorlike fermions which couple
to it. The production and detection possibilities of such singlet
neutral scalars at hadron colliders are considered for different
scenarios of vectorlike fermions.  We find that for some values of
masses and couplings, detection at the CERN Large Hadron Collider (LHC)
appears to be a distinct possibility, while at the Fermilab Tevatron
upgrade the $h^0$ might be observed only in very favourable
circumstances.

{\footnotesize\em Electronic addresses:}
$^\heartsuit$anindya@cubmb.ernet.in;
$^\dagger$amitava@cubmb.ernet.in;\\$^\#$sreerup@theory.tifr.res.in

\newpage
\section{INTRODUCTION}
\rm

Scalar higgs fields are an important ingredient in the Standard Model
(SM) and its popular extensions. Searches for higgs bosons are,
therefore, among the important objectives at present and projected
colliders.  Recent measurements of the $\rho$ parameter (or the oblique
$T$ parameter) at the CERN $e^+ e^-$ collider LEP-1 yield a result very
close to unity \cite{PDG}, which limits the natural possibilities for
light scalars to just singlets and doublets of the $SU(2)_L$ component
of the SM gauge group. The SM adopts a one-doublet scenario,
while extra doublets appear in many of its extensions including the
minimal supersymmetric standard model (MSSM) and the left-right
symmetric model.  Phenomenological consequences of one or more doublets
and their detection possibilities have been extensively studied in the
literature \cite{HiggsHunter,Kunszt}, but the singlet option has not
received the same kind of attention. Save for some discussion in the
context of Majoron models and non-minimal supersymmetric models
\cite{HiggsHunter} one comes across very few studies of singlet higgs
bosons.

Interestingly, neutral $SU(2)_L$ singlet scalar particles are predicted
in several extensions of the SM. They are present in a natural manner
in many Grand Unified Theories (GUTs). For example, the fundamental
{\bf 27}-plet of $E_6$, utilised for spontaneous symmetry breaking,
includes several such fields \cite{E6}. The next to minimal
supersymmetric standard model (NMSSM) \cite{NMSSM} has just such an
extra field to generate the higgs mass parameter -- the so-called $\mu
$-term.  Other models -- {\it e.g.}, the left-right symmetric
$SU(2)_L\times SU(2)_R \times U(1)$ model and its GUT extensions like
$SO(10)$ \cite{SO(10)} -- also include singlet scalars. It has also
been stressed \cite{Pati} that in a class of composite models of quarks
and leptons such a neutral singlet scalar is an essential prediction.
Moreover, it has been shown \cite{Kundu} that the addition of a singlet
higgs scalar (with or without an extra generation of vectorlike
fermions) can provide a realistic solution to the fine-tuning problem
in the SM. It is therefore of interest to consider strategies for the
detection of these scalar singlets (henceforth called singlet higgs
bosons) at the present and upcoming colliders.

Barring mixing with the doublet higgs bosons \cite{Binoth}, the singlet
scalars will not couple to ordinary quarks and leptons. Rather, they
will couple to vectorlike fermions -- quarks and leptons whose left-
and right-handed components transform identically ({\em i.e.}, both
singlets or both doublets) under $SU(2)_L$.  Many of the models with
singlet higgs bosons include such quarks and leptons.  Prominent among
these are the $E_6$ GUT models which contain vector singlet quarks of
charge $-\frac{1}{3}$ as well as vector singlet and vector doublet
leptons. The composite models of Ref. \cite{Pati} also contain
vectorlike fermions which play an important role in explaining the
masses and mixings of the usual quarks and leptons.

The singlet higgs boson could be produced at the CERN $e^+e^-$ collider
LEP-1 through the vectorlike fermion loop induced decay $Z^0
\rightarrow h^0 \gamma$. In an earlier paper \cite{Raychaudhuri} two of
the authors have considered, with reference to the $e^+e^-$ collider
LEP-1, models in which a real singlet higgs boson $h^0$ occurs together
with vectorlike quarks and leptons.  Unfortunately, though the signal
is relatively clean, the number of such events generally turns out to
be too small for effective detection at LEP-1 --- even with a catch of
$10^7 Z^0$s.

In the present work we analyse possibilities for producing a singlet
higgs boson $h^0$ from gauge boson fusion at hadron colliders.  The
$h^0 g g$ (or $h^0 \gamma \gamma$, $h^0 Z^0
\gamma$, $h^0 Z^0 Z^0$, $h^0 W^+ W^-$) interaction will be mediated by
a triangle diagram containing  vectorlike quarks (leptons) just as the
SM $H^0 gg$ interaction is mediated by a top-quark triangle. In fact,
as the masses and couplings will be chosen to be rather similar, the
numbers produced in the two processes are comparable. Detection of the
$h^0$, however, will require modified strategies since its decay modes
are quite different from the SM $H^0$ and depend on the vectorlike
fermion scenario being considered. In general, we find the $h^0
\rightarrow \gamma \gamma$ mode to be the most promising one, since the
occurrence of hadronically quiet hard photon pairs with a peak in the
invariant mass is a clear signal for the decay $h^0
\rightarrow \gamma \gamma$. For sufficiently large values of the
singlet mass $m_h$, the $h^0 \rightarrow Z^0 \gamma \rightarrow
\ell^+\ell^- \gamma$ and $h^0 \rightarrow Z^0Z^0 \rightarrow
\ell^+\ell^- \ell^+\ell^-$ modes ($\ell = e,\mu$) also become viable.
We have estimated SM backgrounds to these processes using a parton-level
Monte Carlo event generator and discussed ways of reducing them through
appropriate kinematic cuts.

The plan of this article is as follows. In Section 2 we discuss those
couplings of the singlet $h^0$ at tree level and at one-loop level
which are relevant for this analysis.  We then consider various
possible vectorlike fermion scenarios and discuss their relative
viability insofar as detection of the $h^0$ is concerned. In Section 3
we discuss the possible modes of production of the $h^0$ at hadron
colliders.  Section 4 is devoted to a study of the decay modes of the
$h^0$ in various scenarios and the possible signals. Backgrounds to
these are also analysed in Section 4 and our conclusions are stated in
Section 5.

\bigskip

\section{COUPLINGS AND GENERAL STRATEGY}

As explained above, the neutral singlet scalar $h^0$ has no $SU(2)_L
\times U(1)_Y$ quantum numbers at all and hence does not couple to the
SM gauge bosons at tree level. Apart from quartic interactions of the
form $H^\dagger H h^2$ it has no interactions with the standard model
quarks, leptons and higgs.  However, it can couple to vectorlike
singlet or doublet fermions, and the couplings can be written as
\begin{eqnarray}
{\rm singlet:} &  ~{\cal L}_s^{hf\bar f} & = \bar f_s (\xi_s + i \eta_s
\gamma_5) f_s h \nonumber \\ {\rm doublet:} & ~{\cal L}_d^{hf\bar f} &
= \bar f_d (\xi_d + i \eta_d \gamma_5) f_d h
\end{eqnarray}
where the subscripts $s,d$ refer to singlet and doublet respectively
and $(\overline{f}_d = [\overline{U}_{d}, \overline{D}_{d}])$.  The
vectorlike fermions couple to the $Z^0$ boson as
\begin{eqnarray}
{\rm singlet:} & ~{\cal L}_s^{Zf\bar f} & = -{\frac{g}
{\cos\theta_{W}}}Q_{f} \sin^{2}\theta_{W}
\overline{f_s} \gamma_{\mu} f_s Z^{\mu} \nonumber\\
{\rm doublet:} & ~{\cal L}_d^{Zf\bar f} & = -{\frac{g}
{\cos\theta_{W}}}(Q_{f} \sin^ {2}\theta_{W} - T_{3f}) \overline{f_d}
\gamma_{\mu} f_d Z^{\mu}
\end{eqnarray}
and to photons and gluons through the usual QED and QCD couplings. In
the subsequent discussion the symbol $f$ wil be used to denote
vectorlike fermions generically.  The Yukawa couplings $\xi_{s,d}$ and
$\eta_{s,d}$ are arbitrary and we examine the cases $ \xi = \eta = 1,
\xi =1, \eta = 0$ and $\xi = 0, \eta = 1$.  One notes that the singlet
and doublet vectorlike fermions can have gauge-invariant mass terms and
hence it is not essential to relate the Yukawa couplings to their
masses. One of the results of this is that we have no handle on the
mass of the vectorlike fermions except the lower bound $m_f
>\frac{1}{2} m_Z$ from the non-observation of the decays $Z^0
\rightarrow f \bar f$ at LEP-1. It should be noted that we assume {\it
no mixing} between the singlet higgs boson and its standard counterpart
and similarly between the vectorlike fermions and their ordinary
counterparts. As a result our analysis is not constrained by mass
bounds derived using such mixings.

Since the singlet higgs boson does not couple to any of the
constituents of the proton, it is clear that it cannot be produced in
$pp$ or $p\bar p$ collisions through tree-level diagrams. At the
one-loop level, however, $h^0$ can be created via $gg$ (or $\gamma
\gamma, Z^0\gamma, Z^0 Z^0, W^+ W^-$) fusion through a triangle of
vectorlike quarks or leptons as the case may be. The mechanism is
illustrated in Fig. 1(a) where $V_i,V_j$ can be any of the pairs of
vector bosons mentioned above. Fig. 1(b) exhibits the diagram
corresponding to loop-induced decay of the $h^0$.

To perform a general analysis of the production and decay of the
singlet higgs boson we need to know the partial width ($\Gamma$) and
the branching fractions of processes of the type $h^0 \rightarrow V_i
V_j$.  The fermion in the loop can be any of $f = U_s, D_s, U_d, D_d,
L_s$, $N_d, L_d$, depending on the scenario under consideration, but
not $N_s$.  Here $U, D$ refer respectively to quarks of charge
$\frac{2}{3}$ and $-\frac{1}{3}$ and $L, N$ to leptons of charge -1 and
0.  We can write the transition amplitude for the generic
process as
\begin{equation}
{\cal M} =
\epsilon_\mu^{(i)}(p_i) \epsilon_\nu^{(j)}(p_j) \Gamma_f^{\mu\nu}
\label{HVVCOUPLING1}
\end{equation}
where the effective coupling $\Gamma_f^{\mu\nu}$ can be written in the
schematic form \cite{Raychaudhuri}
\begin{equation}
\Gamma_f^{\mu\nu} = \sum \frac{\alpha \omega_{if} \omega_{jf} m_f}
{\pi} {C_{ij}^f}
\left[ \xi_f (F^{ij}_1 g^{\mu\nu} - F^{ij}_2 p_j^\mu p_i^\nu) +
i \eta_f F^{ij}_3 \epsilon^{\mu\nu\rho\delta} p_{i\rho} p_{j\delta}
\right].
\label{HVVCOUPLING2}
\end{equation}
Here the summation is operative only for the vector doublet fermion
scenario and runs over the members of the multiplet. Further, a sum
over repeated greek indices is implied. The factors $\omega_{if},
\omega_{jf}$ depend on the gauge bosons $V_i,V_j$ and the vectorlike
fermion scenario under consideration.  A list of the possibilities is
given in Table 1. The colour factors $C^f_{ij}$ are given in Table 2.
The presence of an overall $m_f$ can be explained by helicity flip
arguments.  The form factors $F^{ij}_1,F^{ij}_2,F^{ij}_3$ are
calculated in terms of the well-known two- and three-point functions of
't Hooft and Veltman and Passarino and Veltman \cite{Veltman},
\begin{eqnarray}
F^{ij}_1 & = & B_0(m_f,m_f;m_h) - 4C_{24} - \frac{1}{2} m_h^2 C_0
\nonumber \\ F^{ij}_2 & = & 4(C_{23} - C_{22}) - C_{0} \nonumber \\
F^{ij}_3 & = &  - C_{0}.
\label{FORMFACTORS}
\end{eqnarray}
where each of the $C$ functions has arguments
$C(m_f,m_f,m_f;m_{V_i},m_{V_j},m_h)$.  If at least one of the gauge
bosons $V_i,V_j$ is massless, these can be written in closed form
\cite{Raychaudhuri,BargerPhillips}, but if both are massive, they have
to be expressed in terms of rather complicated formulae involving
dilogarithms.  These are evaluated using a computer code
\cite{Mukhopadhyaya} developed using the algorithms of Ref.
\cite{Veltman}.

If $V_i,V_j$ are either photons or gluons, gauge invariance demands
\begin{equation}
F^{ij}_1 =  p_i.p_j F^{ij}_2.
\label{GAUGEINVARIANCE}
\end{equation}
This is explicitly verified by using relations among the $B$ and $C$
functions which obtain when one of the external masses vanish. These
relations can be found in Refs. \cite{Raychaudhuri,Veltman}.

At this juncture it seems appropriate to discuss the vectorlike fermion
scenario(s) considered in this paper and the various search strategies
prescribed for each. Rather than include a complete extra generation of
vectorlike quarks and leptons, we have chosen to consider the different
possibilities one at a time. This has the advantage of simplicity and
is theoretically quite legitimate since  these extra fermion
representations are anomaly-free due to their vectorlike nature.  We
thus have the following options:
\begin{enumerate}
\item
a vectorlike doublet of quarks $\overline{f_d} = (\overline{U_d},
\overline{D_d})$;
\item
a vectorlike singlet quark $f = U_{s}$ of charge $\frac{2}{3}$;
\item
a vectorlike singlet quark $f = D_{s}$ of charge $- \frac{1}{3}$;
\item
a vectorlike doublet of leptons $\overline{f_d} = (\overline{N_d},
\overline{L_d})$;
\item
a vectorlike singlet lepton $f = L_{s}$ of charge $-1$;
\end{enumerate}
A sixth possibility, that of a vectorlike singlet neutrino, has been
discounted because it does not couple to any of the SM gauge bosons.

Let us consider scenario 1 in more detail. Since vectorlike doublet
quarks couple to all the SM gauge bosons, the possibilities in Fig.
1(a) are for $V_i V_j$ to be any of the pairs $gg$, $\gamma\gamma$,
$Z^0\gamma$, $Z^0Z^0$, $W^+W^-$.  Of all these, the only one worth
considering is the $gg$ mode, not only because the $h^0gg$ coupling is
the largest, but because of the high gluon luminosity at a hadron
collider. The production mechanism would be analogous to that envisaged
for the SM higgs boson from gluon-gluon fusion through a top quark
triangle \cite{BargerPhillips}, and the numbers obtained are, in fact,
comparable.  Some of the advantage is lost, however, when we look for
the detectable signals for the $h^0$.  The dominant decay modes of
$h^0$ will be $h^0
\rightarrow gg$ or $h^0 \rightarrow \overline {U_d} U_d, \overline
{D_d} D_d$ depending on the masses of $h^0$ and $U_d,D_d$ (doublet
fermions have to be more-or-less degenerate, from the bounds arising
from the oblique $T$ parameter).  In either case, one would see a pair
of hadronic jets which would be lost in the large QCD background.
Accordingly, we have to turn to the electroweak modes, {\em viz.} $h^0
\rightarrow \gamma
\gamma, h^0 \rightarrow  Z^0\gamma, h^0 \rightarrow Z^0Z^0, h^0
\rightarrow W^+W^-$.  The last possibility, though it has a large
branching fraction (see below), will not be considered further in this
work because the hadronic decays of the $W^\pm$ pair would be swamped
by the QCD background while the leptonic decays would involve missing
transverse energy and momentum due to {\it two} neutrinos, rendering
the analysis based on reconstruction of invariant masses
impossible.  We choose, therefore, to restrict ourselves to the
possibilities that $V_i^\prime, V_j^\prime$ are $\gamma$ or $Z^0$, as
the case may be.  Furthermore we assume that the $Z^0$ is identified by
its charged leptonic decay modes $Z^0 \rightarrow \mu^+\mu^-, e^+e^-$,
(these decay channels contribute 6.6\% to the total $Z^0$ width)
generically denoted $Z^0 \rightarrow \ell^+\ell^-$ in the subsequent
analysis.

Scenarios 2 and 3 are rather similar. However, all electroweak
amplitudes in scenario 3 would be suppressed by the charge
$-\frac{1}{3}$ of the $D_s$ quark (see Table 1) which makes detection
more difficult.

Scenario 4 is interesting since gluonic couplings are disallowed and
one has to rely on electroweak production modes, {\em i.e.} $V_i, V_j$
can be $\gamma
\gamma, Z^0\gamma, Z^0Z^0, W^+W^-$. The lesser numbers of $h^0$s
produced are partially compensated by the large branching ratios now
available for the $\gamma\gamma, Z^0\gamma, Z^0Z^0$ decay modes. Of
course, the tree-level decay $h^0 \rightarrow L^+ L^-$ will be the
dominant one, {\it if allowed kinematically}.  In this case, the heavy
lepton $L^\pm$ should behave rather like a muon, except that its track
will show little or no curvature.  This last circumstance will,
however, render measurement of momenta difficult, so this mode may not,
after all, be a good option to pin down the $h^0$.
Scenario 5 is almost identical to 4 except that the $h^0 \rightarrow
W^+W^-$ mode is absent and the couplings are rather different.

Here it might not be irrelevant to  compare the total width of this
singlet scalar with that of the SM higgs. For the sake of this
discussion we consider two values of the vectorlike fermion mass: 50
and 150 GeV.  In Table 3, we present the widths of  $h^0$ in the
various scenarios for $\xi = \eta = 1$.  For comparison, the width of
the `Standard' Model higgs for a hypothetical top quark mass of 50 and
150 GeV are also presented. Note that beyond the 2$m_f$ threshold the
singlet higgs width is large as a consequence of the tree level decay
and the choice $\xi=\eta = 1$.

As noted earlier, the signal for production and decay of an $h^0$ at a
hadron collider is a pair of vector bosons $V_j^\prime, V_j^\prime$,
which are either of $\gamma$ or $Z^0$, and whose invariant mass shows a
sharp peak (below the $m_h = 2 m_f$ threshold) at the singlet higgs
mass. The principal SM background to each of these processes comes from
$q \overline {q}$ annihilation and gluon gluon fusion  to a pair of
vector bosons $V_i^\prime, V_j^\prime$ through a box diagram. (For the
$\gamma \gamma$ final state, {\it bremsstrahlung} is also an important
background.) These backgrounds could be quite significant even after
imposing various kinematic cuts. The search strategy suggested,
therefore, is to plot the invariant mass distribution of the final
decay products in suitable bins. Presence, or otherwise, of an $h^0$
component will be indicated by an excess of events in the particular
bin where the peak of the signal lies (this will naturally depend on
the mass of the $h^0$). The subsequent discussions are based on this
strategy.  For the decay $h^0 \rightarrow Z^0\gamma$, since we identify
the $Z^0$ by its decay to $\ell^+\ell^-$, another possible background
will be the radiative process $q \overline {q} \rightarrow  \ell^+
\ell^- \gamma$.  This last background can, however, be easily dealt
with by requiring isolation of the photon from the $l^+$ and $l^-$ and
demanding that the invariant mass of the lepton pair be around
$m_Z$.

\bigskip

\section{HADRONIC PRODUCTION OF THE $h^0$}
\normalsize\rm

It has already been mentioned above that the mechanism for producing
$h^0$ in $pp$ or $p \bar p$ collisions will depend upon the  vectorlike
fermions in the loop. So long as these are quarks, the dominant process
will be through $gg$ fusion via a vectorlike quark ($Q$) triangle, the
effective $h^0gg$ coupling being given by Eqs. (\ref{HVVCOUPLING1}) ---
(\ref{GAUGEINVARIANCE}). Using these, we obtain, at the parton level,
the cross section
\begin{eqnarray}
\hat \sigma (gg \rightarrow h^0) & = &
\frac {\alpha _{s}^{2}m_{Q}^{2}} {64 \pi} \sqrt {\hat s}
\big( \xi_Q^2 \mid F^{gg}_2 (\hat s)\mid^2 +  \eta_Q^2 \mid F^{gg}_3
(\hat s)\mid^2 \big)
\delta (m_{h} - \sqrt {\hat s}) \nonumber \\
& = & \frac{\pi^2}{16\hat s} \Gamma (h^{0}\rightarrow gg)
\delta (m_{h} - \sqrt {\hat s})
\end{eqnarray}
where $\sqrt{\hat s}$ is the total centre-of-mass energy of the
colliding partons and $\Gamma (h^0 \rightarrow gg)$ is the decay width
of an $h^0$ to a $gg$ pair.  In view of the other uncertainties
involved as regards masses and couplings, we have not included QCD
corrections -- which can be fairly substantial \cite{Spira} -- to this
process.  To obtain the inclusive hadronic cross section for $h^0$
production, $\hat \sigma (gg \rightarrow h^0)$ must be convoluted with
the distribution functions $f_{p/g}(x)$ for the gluons. The resulting
formula (for scenario 1) is
\begin{equation}
\sigma (pp, p \overline {p} \rightarrow h^0 + X)
= \frac{\pi^2}{8sm_h} \Gamma (h^{0}\rightarrow gg)
\int_\tau^1 {dx \over x} f_{p/g}(x) f_{p/g} ({\tau \over x})
\end{equation}
where $\sqrt{s}$ is the centre-of-mass energy of the colliding hadrons
and $\tau$ is a dimensionless quantity given by $m^2_h/s$.

To obtain numerical estimates for the production cross section we have
used a parton-level Monte Carlo event generator incorporating the
recent structure function parametrisations of Martin, Roberts and
Stirling
\cite {MRS}. The resulting estimates for production of $h^0$ for three
different choices of $\xi$ and $\eta$ are illustrated in Fig. 2 (a),
(b) and (c) for a vectorlike doublet of quarks (scenario 1). The solid
curves correspond to  $\sqrt{s} = 14$ TeV (LHC) and the dashed curves
correspond to $\sqrt{s} = 1.8$ TeV (Tevatron). The kink near $m_h =$
100 (400) GeV in each of these curves corresponds to the $m_h = 2m_Q$
threshold for $m_Q =$ 50 (200) GeV where the numerical results are not
very reliable.

In Fig. 2, for $\sqrt{s} = 14$ TeV, the cross section has been
multiplied by a luminosity of ${\cal L} = 10^5$ pb$^{-1}$/year ({\em
i.e.} the so-called high luminosity option at the LHC) while for
$\sqrt{s} = 1.8$ TeV, the corresponding luminosity has been taken to be
${\cal L} = 10^3$ pb$^{-1}$/year {\em i.e.} for the projected
Tevatron*. It may be seen that at the Tevatron*, one could produce over
$10^4$ $h^0$s per year for $m_h < 200$ GeV for $m_Q = 50$ GeV; this
number drops to 10 when $m_h$ becomes 500 GeV.  For $m_Q = 200$ GeV,
the number of $h^0$s produced per year is less than the number obtained
with $m_Q = 50$ GeV over the entire mass range of $h^0$ except when
$m_h$ crosses the $2m_Q$ threshold, when the numbers obtained with the
two different fermion masses are more or less the same. This is
inevitable in view of the fact that the $h^0gg$ coupling falls rapidly
with $m_Q$. It is interesting to note the contrast with the case of the
SM $H^0gg$ coupling, which becomes roughly constant for large $m_t$.
This is because the $H^0t\bar t$ coupling  is proportional to $m_t$
whereas we have taken the $h^0Q\overline {Q}$ coupling  to be a
constant (see above).

At the LHC, with the high luminosity option, we immediately notice that
one could produce over $10^8$ $h^0$s per year for $m_h < 200 $ GeV with
$m_Q = 50 $ GeV.  The corresponding numbers for $m_Q = 200$ GeV are
about an order of magnitude smaller.

We see from the Figs. 2(a), (b), and (c) that the number of $h^0$s
produced for three different choices of $\xi$ and $\eta$ do not produce
any significantly different result. Also the trend of variation with
$m_h$ is more or less the same in the above three cases. Thus in our
analysis we shall use $ \xi = \eta = 1$ from now on.

The corresponding estimates for scenarios 2 and 3 are obtained in the
same way and are smaller than in scenario 1 as there is just one quark
-- singlet -- in the loop. Fig. 2(d) shows the  number of $h^0$s
produced for singlet $U$-type quarks at the LHC and the Tevatron.

The situations for scenarios 4 and 5 are quite different. Since the
loop-fermion is now a lepton, there is no $h^0gg$ coupling. Accordingly
we should now expect the $h^0$ to be produced from the fusion of
$\gamma\gamma, \gamma Z^0, Z^0Z^0$ (and in scenario 4 from $W^+W^-$)
pairs emitted from the parent hadrons. Such electroweak production of
the $h^0$ naturally leads to lower cross sections, not only because of
the small couplings involved in emission of $\gamma, Z^0$ from the
parent quarks, but also due to the lower flux of quarks coming from the
proton compared with that of gluons at high energies, especially at the
LHC.  The parton-level cross section for $\gamma
\gamma \rightarrow h^0$ is easily obtained by multiplying the same for
the $gg \rightarrow h^0$ case by the factor $(e \omega_{\gamma f} /g_s
\omega_{gf})^4$.  The values of $\omega_{\gamma f}$ and $\omega_{gf}$
are given in Table 1 and are different for the cases 4 and 5. As in the
case of two-gluon fusion, in order to obtain the cross section at the
hadron level, one must convolute the parton-level cross section with
the quark distribution in the proton (or antiproton) as well as the
probability for the emission of a photon from a quark. The relevant
formulae can be obtained using the well-known effective photon
approximation and are given in Ref.
\cite{BargerPhillips}. Accordingly, we have
\begin{eqnarray}
&&\sigma(pp, p \overline {p} \rightarrow h^{0} + X)  \\ && =
\frac{8 \pi^2}{s m_h} \Gamma (h^0 \rightarrow \gamma \gamma)
\nonumber \\
&& \times \sum_{a,b} \int^1_\tau dx_1 \int^1_{\tau/x_1} dx_2
\int^1_{\tau/x_1 x_2} dx_3 \frac{1}{x_1 x_2 x_3} f_{q_a/\gamma}(x_1)
f_{q_b/\gamma}(x_2) f_{p/q_a}(x_3) f_{p/q_b}(\frac{\tau}{x_1 x_2 x_3})
\nonumber
\end{eqnarray}
where the sum over $a,b$ runs over valence quarks only. When the $h^0$
is produced from the fusion of the massive vector bosons $W^\pm, Z^0$,
the calculation becomes more complicated. The probability of emission
of a massive vector boson from a quark depend on its polarisation, and
hence one must consider the polarised amplitudes ${\cal M}_{\lambda_1
\lambda_2}$ rather than the spin-averaged one we have been discussing
till now. The nonvanishing amplitudes are
\begin{eqnarray}
{\cal M}_{00} & = &
\frac{\alpha m_f \omega_{if} \omega_{jf} \xi_f} {\pi m_{V_i} m_{V_j} }
\Bigg[ \frac{1}{2}(m_h^2 - m_{V_i}^2 - m_{V_j}^2 ) F^{ij}_1
+ \frac{1}{4} \lambda(m_h^2, m_{V_i}^2, m_{V_j}^2) F^{ij}_2 \Bigg]
\nonumber \\ {\cal M}_{+-} & = & - \frac{\alpha m_f \omega_{if}
\omega_{jf}}{\pi}
\Bigg[ \xi_f F^{ij}_1 - \frac{\eta_f}{2}
\sqrt{\lambda(m_h^2,m_{V_i}^2, m_{V_j}^2)} F^{ij}_3 \Bigg] \nonumber \\
{\cal M}_{-+} & = & - \frac{\alpha m_f \omega_{if} \omega_{jf}}{\pi}
\Bigg[ \xi_f F^{ij}_1 + \frac{\eta_f}{2}
\sqrt{\lambda(m_h^2,m_{V_i}^2, m_{V_j}^2)} F^{ij}_3 \Bigg]
\end{eqnarray}
where $\lambda(x,y,z) \equiv x^2 + y^2 + z^2 - 2xy - 2yz - 2zx$ and
$i,j$ are either $W,W$ or $Z^0,Z^0$ or $Z^0, \gamma$.  The
polarisation-dependent probabilities of emission of massive
vector-bosons from a parton  are calculated in the effective $W,Z^0$
approximations for $\sqrt{s} \gg m_{W,Z}$ and are given in Ref.
\cite{Dawson}. The final formulae are analogous to Eq. (9) but are
rather cumbersome and have not been presented explicitly.

Numerical estimates for the electroweak production of $h^0$s may be
obtained from a parton-level Monte Carlo event generator as before, and
are presented in Fig. 2(e) for scenario 4. The numbers are rather small
and the behaviour with growing $m_h$ more or less mimics that in the
case of strong production.  At the Tevatron, $h_0$ can be produced only
via photon photon fusion. But at the LHC it can also be produced via
$Z^0 Z^0$, $W^+ W^-$ or $ \gamma Z^0$ fusion. The sharp peaks in the
plots correspond to the thresholds at $ m_h = m_Z, 2m_Z ,2m_W$ and as
usual at $2m_f$.  Observing the smallness of the numbers of $h_0$
produced, we will not further analyse the signals from this scenario
(and scenario 5, which is similar).
\bigskip

\section{DETECTION POSSIBILITIES}
\normalsize\rm

Once produced, the $h^0$ will decay to a pair of vector bosons
$V_i^\prime, V_j^\prime$ through a vectorlike fermion triangle, or, in
case it is sufficiently massive, to a pair of vectorlike fermions. If
the vectorlike fermions are quarks ($Q$), then the dominant decay modes
will be $h^0 \rightarrow gg$ for $m_h < 2m_Q$ and $h^0 \rightarrow Q
\overline {Q}$ for $m_h > 2m_Q$.  Neither of these decays will be
observable, however, because of the large QCD background. We turn,
therefore, to the electroweak decay modes.

The branching ratios for the various decay channels are plotted as
functions of $m_h$ (for $m_f$ = 100 GeV) in Figs. 3(a) and  (b) for the
vectorlike fermion scenarios  1 and 2 respectively. The convention
followed regarding the different curves is:
\begin{enumerate}
\item  solid: $h^0 \rightarrow \gamma \gamma$ mode;
\item  solid with dots: $h^0 \rightarrow Z^0 \gamma$ mode;
\item  large dashes: $h^0 \rightarrow Z^0 Z^0$ mode;
\item  dot-dash: $h^0 \rightarrow W^+W^-$ mode;
\item  small dash: $h^0 \rightarrow f \overline {f}$ mode;
\item  dots:  $h^0 \rightarrow gg$ mode.
\end{enumerate}
As one would expect, there is no $W^+W^-$ mode in Fig. 3(b).

It has already been explained that the $W^+W^-$ decay mode in scenario
1, though quite prominent, may not be viable for detection due to the
missing energy carried away by two neutrinos, so we shall concentrate
on the $\gamma\gamma, Z^0 \gamma , Z^0Z^0$ modes only.  For scenarios 1
and 2 (and also 3, though this is not shown), when the vectorlike
fermion are quarks, these branching ratios are rather small, being of
the order of $10^{-3}$ or less. However, in these three cases, many
more $h^0$s are produced because of the gluonic mode of production, so
this disadvantage is more than offset as we shall see presently. One
notes that for singlet fermions, the couplings of $\gamma$ and $Z^0$
are proportional to the charge of the fermion, as a result of which
(and the favourable kinematics) the $\gamma\gamma$ mode is the dominant
one of the three.  For vectorlike doublets, however, the $Z^0$ couples
more strongly than the photons, so that modes with a final $Z^0$ are
enhanced above the purely photonic mode.  In fact, the $Z^0Z^0$ mode,
though suppressed by a factor of $\frac {1}{2}$ compared with the
$Z^0\gamma$ mode because of Bose statistics, still turns out to be
dominant because of the presence of a vectorlike neutrino triangle in
the doublet case.  However, since we consider the detection of the
$Z^0$ through its $\ell^+\ell^-$ decay channel, both signal and
background are suppressed by the relevant branching ratio.

Since the mass of the singlet higgs boson $h^0$ is unknown, it is
convenient to divide the possible mass range into four regions,
somewhat as is done in the case of the SM higgs boson. These are

\begin{enumerate}
\item  Very light: $m_h < 50$ GeV;
\item  Light: 50 GeV $< m_h < m_Z$;
\item  Intermediate mass: $m_Z < m_h < 2m_Z$;
\item  Heavy:  $m_h > 2m_Z$.
\end{enumerate}

For light and very light singlet higgs bosons, the only viable decay
mode is $h^0 \rightarrow \gamma \gamma$; for intermediate mass, the
decay $h^0
\rightarrow Z^0 \gamma$ becomes available; while for the heavy case we
also have the decay $h^0 \rightarrow Z^0Z^0$. Assuming the $Z^0$ is to
be identified by its $\ell^+\ell^-$ decay mode, we accordingly look for
either $\gamma \gamma$, or $\ell^+ \ell^-\gamma$ or the so-called `gold
plated' signal $\ell^+ \ell^- \ell^+ \ell^-$ respectively.  Throughout
the mass range of the $h^0$ the $h^0 \rightarrow
\gamma \gamma$ mode remains viable and the role of the other
modes is to present further options which can add to the signal. For
high values of $m_h$, $h^0$ production itself goes down, as shown in
Fig. 2, and it becomes desirable to look for the the singlet higgs in
as many channels as possible.

As we have stated earlier, the principal SM background to each of these
processes will come from the tree-level process $q \bar q \rightarrow
V_i^\prime V_j^\prime$ and gluon gluon fusion through a box diagram to
$V_i^\prime V_j^\prime$. ({\it Bremsstrahlung} makes an
additional important contribution to the two photon background.) We
evaluated the tree diagram contribution using a parton-level Monte
Carlo event generator and multiplied the numbers by appropriate
factors to take the other process(es) into account (see later). In
general, for these processes the vector bosons will be produced closer
to the beam-pipe and with softer transverse momentum distributions than
in the case of the signal, so that kinematic cuts are helpful to reduce
the backgrounds.  Even with the above cuts, the signal is quite often
smaller than the background, decreasing, in fact, as $m_h$ increases
because of a fall in the number of $h^0$s produced.  Accordingly, we
adopt the strategy suggested in sec. 2, {\em viz.} we plot the
distribution in the invariant mass of the final products $\gamma
\gamma$ or $\ell^+ \ell^-\gamma$ or $\ell^+ \ell^- \ell^+
\ell^-$, as the case may be, and look for a peak indicative of the
decay of an $h^0$.

We shall now turn to the results obtained with a parton-level Monte
Carlo generator for the above signals and their respective backgrounds.
In each case, we present numbers for the Tevatron* and the LHC (high
luminosity option) separately, for the different vectorlike fermion
scenarios enumerated above. It may be noted at the very outset that we
have used the {\em same} kinematic cuts (see above) for all the
vectorlike fermion scenarios considered and also for studies at the
Tevatron with $\sqrt{s} = 1.8$ TeV and the LHC with $\sqrt{s} = 14$
TeV. This rather artificial choice is purely for purposes of comparison
and is not suggested as a prescription for a realistic analysis. One
can, for example, relax the cut $E_T^\gamma > 25$ GeV on the transverse
energy of the final state photons in the case of vectorlike
doublet quarks (scenario 1) at LHC with the high luminosity option, and
thereby obtain some information about very light singlet higgs bosons.
However the analysis presented in this work is meant to be illustrative
and some of the results can be improved if we fix upon a
particular vectorlike fermion scenario. Results at the Tevatron upgrade
and at the low luminosity option of the LHC can be obtained by
scaling the signal by a factor of $\frac{1}{10}$ in either case, while
the $1\sigma$ fluctuations in the background get scaled by
$\frac{1}{\sqrt{10}}$. Since the graphs are plotted on a logarithmic
scale, these factors simply correspond to vertical shifts of the entire
curve(s) and the numbers may be easily read off.

{\em\underline {The $h^0 \rightarrow \gamma \gamma$ mode:}} For 50 GeV
$< m_h < m_Z$ the only viable decay mode is $h^0 \rightarrow \gamma
\gamma$.  This is also the dominant one of the electroweak modes with
{\it singlet} fermions in the loop for all mass ranges.  The signal
will be a pair of hard photons produced back-to-back in the $h^0$ rest
frame (but not in the laboratory frame), whose invariant mass,
$M_{\gamma\gamma}$ has a sharp peak around $M_{\gamma\gamma}=m_h$.  In
Fig. 4(a) we have shown, for scenario 1, {\em i.e.} a doublet of
vectorlike quarks, the number of events expected per year as a function
of $M_{\gamma \gamma}$ -- in bins of 10 GeV -- at the LHC with the high
luminosity option (see above). The solid, large dashed, small dashed
curves represent the expected signal in relevant bin (should the
singlet $h^0$ have a mass which falls in that particular bin) for $m_Q$
= 50, 100, 200 GeV respectively. In view of the sharpness of the
resonance, the entire signal will lie in the relevant bin \cite{Width}.
The kinks correspond to the $m_h = 2m_Q$ thresholds. It may be noted
that this contribution goes down more or less steadily as the invariant
mass increases.  This simply reflects the fall in $h^0$ production for
increasing $m_h$. In this analysis, we have imposed a cut on the photon
pseudo-rapidity $\eta_\gamma < 2.5$.  To obtain a viable signal, it
usually becomes necessary to impose a further cut $E_T^\gamma > 25$
GeV. This tells us that {\em very light singlet higgs bosons (of mass
$m_h < 50$ GeV) are unlikely to be seen at hadron colliders}.

The histogram shows the square root of the number of events from the SM
background deposited in each bin\footnote{This convention will also be
followed in figs. 5-7.} which is a reasonable measure of the 1$\sigma$
fluctuation. We have multiplied the numbers obtained from the $q \bar q
\rightarrow \gamma \gamma$ Monte Carlo by a factor of 8
\cite{GammaBack} to take into account the di-photon production from
gluon gluon fusion and the {\it bremsstrahlung} contribution. It may be
pointed out that the background will be suppressed by the fact that the
$\bar q$ is a sea-quark at the LHC.  Since the signal will be seen as a
peak in a particular bin over and above the SM background (and its
fluctuations), it is clear that the numbers shown in Fig. 4(a) indicate
that detection of the singlet $h^0$ will be viable in the entire range
$m_h=$ 50 --- 400 GeV for $m_Q =$ 200 GeV considered in this paper
since we will get more than a $5\sigma$ peak in the invariant mass
distribution of the photons. With the low luminosity option, it may be
difficult to probe more than $m_h
\simeq 100$ GeV if $m_Q \simeq 50$ GeV, though this limit goes up to
about 200 GeV if $m_Q$ is 100 GeV or more.

We have chosen bins of 10 GeV since this appears typical of present
experiments \cite{CDF}. A coarser resolution will not usually affect
the signal, but will increase the background and its fluctuations.  For
example, if the data is collected in bins of 20 GeV, the background
will increase by a factor of about 2 and its fluctuations by about 1.4.
This will hardly affect numbers in the high luminosity option, but will
reduce the discovery limits for the low luminosity option still further
by about 50 GeV in each case.

Fig. 4(b) shows the same curves for scenario 2, {\em i.e.,} the quark
doublet is now replaced by a singlet $U$-quark. The curves are roughly
similar, except for the slightly smaller number of $h^0$ produced in
this case, and the comments made regarding Fig. 4(a) are equally
applicable to this case.

Figs. 5(a) and (b) illustrate the same numbers for scenarios 1 and 2
respectively at the Tevatron* with a centre-of-mass energy of 1.8 TeV
and a luminosity of $10^3$ pb$^{-1}$. As in the case of Fig. 4, we have
displayed results for $m_Q = 50, 100, 200$ GeV (the solid, large dashed
and small dashed curves respectively) and a histogram representing the
$1\sigma$ fluctuation in the background.  Since the Tevatron is a $p
\bar p$ collider, both the $q$ and the $\bar q$ can be valence quarks.
Hence, there is no suppression of the background as was the case for
the LHC.  This is unfortunate for the kind of signal we are
investigating in this work.  As is clear from Fig. 5, one cannot expect
a reasonable signal at the Tevatron* very much above 100 GeV for the
most promising situation of light vectorlike quarks in the loop ($m_Q =
50$ GeV). For larger values of the quark mass, the situation is much
worse and a $1\sigma$ effect is just obtained.  At the Tevatron with a
luminosity of 100 pb$^{-1}$ one can at best obtain a $2\sigma$ effect
if the masses of both $h^0$ and the vectorlike fermion are in the
vicinity of 50-60 GeV.  Electroweak production at this energy is too
small to yield even a single $h^0$ with the design luminosities, so
the corresponding graphs in scenarios 4 and 5 have not been shown. It
is obvious that data from the Tevatron or Tevatron* are not likely to
impose serious constraints on the scenarios being considered in this
work, except for the corner of parameter space where the masses of the
singlet and the vectorlike fermions are light and their Yukawa
couplings are large.

{\em\underline {The $h^0 \rightarrow Z^0 \gamma$ mode:}} For $m_h >
m_Z$ the $h^0$ can decay into a $Z^0 \gamma$ pair as well as a $\gamma
\gamma$ pair.  Fixing on the $\ell^+\ell^-$ decay mode of the $Z^0$, we
look for an isolated photon and a pair of leptons with the demand that
the total invariant mass $M_{\ell^+\ell^-\gamma}$ of the final state
has a sharp peak which would indicate the presence of an $h^0$
component. To remove the radiative background from $q \bar q
\rightarrow \ell^+\ell^-\gamma$ we also require the invariant mass of
the $\ell^+\ell^-$ pair to lie in the vicinity of $m_Z$.

In Fig. 6(a) we illustrate, as before, the distribution in invariant
mass for the final products at the LHC with the high luminosity option,
subject to the following kinematic cuts: (a) Transverse energy of the
photon is greater than 25 GeV; (b)  Transverse momentum of the leptons
are each greater than 20 GeV; (c)  All decay products have
pseudo-rapidity $\eta < 2.5$; (d)  The photon is isolated from the
lepton, {\em i.e.} $\theta_{\gamma l} > 15^0$; (e)  The invariant mass
of the lepton pair lies between 85 to 95 GeV.  (f) The angle between
the photon and the reconstructed $Z^0$ is greater than $10^0$. (This
helps in removing a significant part of the background.)

The signal is rather small in the range $m_h = 90 -120$ GeV and has not
been exhibited here. As regards the background in this decay channel,
we examine the $q\overline {q} \rightarrow \gamma Z^0 \rightarrow
\gamma l^+ l^-$ and the radiative process $q \overline {q} \rightarrow
\gamma l^+ l^-$ using event generators. The latter contribution is
essentially eliminated by the 85 GeV $ < m_{l^+l^-} < 95 $ GeV cut. To
estimate the contribution to the background from  gluon-gluon fusion,
we multiply the numbers by a factor of 1.3 \cite{ZBack}. For doublet
fermions in the loop, the signal is larger than a $5\sigma$ fluctuation
of the background upto 400 GeV higgs mass with $m_Q$ = 200 GeV. With
$m_Q$ = 50 (100) GeV, the signal is more than the $5\sigma$ fluctuation
of the background upto $m_h$ = 170 (220) GeV.

Fig. 6(b) contains the results for scenario 2 ({\em i.e.} for singlet U
type quark) and the situation is not as promising. Here also we see
that for $m_Q$ = 200 GeV the signal is above a $2\sigma$ to $5\sigma$
fluctuation of background when $m_h$ changes from 150 to 400 GeV. For
$m_Q$ = 100 GeV, below $m_h$ = 200 GeV  the signal is very large
compared to the background but it falls sharply as $m_h$ crosses the $2
m_Q$ threshold.  For $m_Q$ = 50 GeV, the situation is hopeless as the
signal is less than 1$\sigma$ fluctuation of the background for the
entire higgs mass range.  This seemingly paradoxical result, in view of
the propagator effect with increasing $m_Q$, is easily explained by
considering the branching ratios of Fig. 3. The signal drops rapidly
beyond $m_h = 2m_Q$ simply because of the opening-up of the $h^0
\rightarrow Q \overline {Q}$ channel. If one considers the low
luminosity option, the predictions for scenario 1 are rather similar to
the predictions for scenario 2 with the high luminosity option.
Scenario 2 with the low luminosity option, is, however, no longer
detectable.

At the Tevatron the signal remains below a 1$\sigma$ fluctuation of the
background starting from low higgs mass upto higher ones. We do not
present these numbers.

{\em\underline {The $h^0 \rightarrow Z^0Z^0$ mode:}} Finally, we
consider the possibility that  $m_h > 2m_Z$. For this mass range, the
$h^0\rightarrow Z^0Z^0 \rightarrow \ell^+\ell^-\ell^+\ell^-$ channel
becomes available in addition to the ones considered before. One notes
that out of a final state $\ell_1^+\ell_2^-\ell_3^+\ell_4^-$ it is
possible to pair the $\ell^+$s and $\ell^-$s in two ways, {\it viz.}
$\ell_1^+\ell_2^-, \ell_3^+\ell_4^-$ and $\ell_1^+\ell_4^-,
\ell_3^+\ell_2^-$. Only one of these sets corresponds to a process with
$h^0 \rightarrow Z^0Z^0 \rightarrow \ell^+\ell^-\ell^+\ell^-$ and it is
for this pairing that the invariant masses of the two $\ell^+\ell^-$
pairs will peak around the $Z^0$-boson mass.  Demanding, therefore,
that two of the four possible $\ell^+\ell^-$ pairs that can be formed
have invariant masses close to $m_Z$, one can remove most of the
backgrounds except, naturally, those due to $q
\overline {q} \rightarrow Z^0Z^0$. Once again we take into account the
gluon gluon fusion to a pair of $Z^0$s by multiplying the $q
\overline {q} \rightarrow Z^0Z^0$ contribution by a factor of 1.3
\cite{ZBack}.  One then considers the invariant mass of all four
leptons in the final state, more or less as was done in the previous
cases. This is a very clear signal and its analogue for the SM $H^0$ --
where it is a tree-level decay -- is widely referred to as a `gold
plated' signal. Unfortunately, however, the rather large mass of the
$h^0$ leads to production of smaller numbers in the first place, so
that this signal ultimately turns out to be less promising than the
previous ones.

For scenario 1, Fig. 7 shows the distribution in invariant mass
$M_{\ell^+\ell^-\ell^+\ell^-}$ for the four final-state leptons in the
mass range 180 to 500 GeV at the LHC with the high luminosity option.
These plots have been obtained with the kinematic cuts: (a)  Transverse
momentum of each lepton is greater than 20 GeV; (b)  All the leptons
have pseudo-rapidity $\eta < 2.5$; (c)  Two of the four possible
$\ell^+\ell^-$ invariant masses lie in the region 85-95 GeV (see
above).  (d) The angle betweeen the two reconstructed $Z^0$s is greater
than $10^0$.

Once again, these cuts are quite adequate to suppress the background.
Though the signal itself is small, one may nevertheless find evidence
for the singlet higgs in this channel if $m_h< 210$ GeV for $m_Q$ = 50
or 100 GeV while for $m_Q$ ($\simeq 200$ GeV) an upper limit of around
400 GeV can be probed.  The situation becomes much worse with the low
luminosity option when only the range $m_h = 180-200$ GeV can yield an
acceptable signal, and that too for $m_Q \simeq 50$ GeV only.  At the
Tevatron upgrade and the Tevatron*, one hardly predicts anything
observable except in the narrow band $180-185$ GeV, which is not worth
investigating in this mode.

{}From the above discussion, and the illustrative numbers presented, it
appears that at the LHC, with high luminosity, there is a good chance
that one will see a signal for the $h^0$, especially if it
appears in conjunction with a doublet of vectorlike  quarks. The
question that immediately comes to mind is: how can one distinguish it
from the SM higgs boson, which has similar decay modes. To answer this,
one must recollect the fact that the SM $H^0$ has a tree-level coupling
to $b\overline {b}$ as a result of which its branching ratios to vector
boson pairs (except to $W^+W^-$ and $Z^0Z^0$, which also occur at tree
level) are strongly suppressed.  Of course, if a tiny signal,
compatible with the SM, should be observed, it could equally well be
due to an $h^0$, with the amplitudes suppressed by small values of
$\eta_f$ and $\xi_f$ or large values of $m_f$.  The distinguishing
feature will be the absence of tree-level $b \overline {b}, Z^0Z^0$ and
$W^+W^-$  decays. Thus, a higgs boson signal through the processes $H^0
\rightarrow \gamma\gamma, H^0 \rightarrow Z^0\gamma, H^0
\rightarrow Z^0Z^0$, but unaccompanied by the other signatures for the
SM $H^0$, could very well be a signal for a singlet $h^0$.

\bigskip

\section{CONCLUSIONS}
\normalsize\rm

We have investigated possibilities for the detection of $SU(2)_L\times
U(1)_Y$ singlet scalars, at the LHC and the Tevatron. The production
and decay modes of these scalars depend on the presence of  vectorlike
fermions whose left- and right-handed components transform identically
under the gauge group. Setting aside the $h^0 \rightarrow gg$ decay
mode, which is expected to be swamped by QCD backgrounds, the
detectable decays are $h^0 \rightarrow \gamma\gamma, h^0 \rightarrow
Z^0\gamma$, and $h^0 \rightarrow Z^0Z^0$ where the $Z^0$ subsequently
decays to a pair of electrons or muons. Among these, the first process
$h^0\rightarrow \gamma\gamma$ seems to be the most promising. The other
two channels can be interesting if the singlet scalar comes in
conjunction with a vectorlike doublet of quarks, but not otherwise.

In all the channels that we have examined there is a sharp drop in the
signal above the $m_h = 2 m_Q$ threshold due to the opening up of the
tree-level $h^0 \rightarrow Q \overline{Q}$ decay mode. Thus for a
vectorlike fermion mass of 50 GeV only a rather limited region of
singlet higgs mass can be explored. On the other hand, for $m_Q$ = 200
GeV the signal remains above the 5$\sigma $ fluctuations of the
background upto large higgs masses. It is easily seen that for larger
$m_Q$, though the higgs mass threshold is increased, the signal itself
is reduced and will not be more than the 5$\sigma $ fluctuation of the
background.

Irrespective of the decay mode, our findings indicate that it is
unlikely that an $h^0$ signal will be seen at the Tevatron.  Only with
the commissioning of the LHC can we look forward to a potential
detection of this particle.  A very light singlet $h^0$ with mass less
than 50 GeV will escape even these tests and one will have to look for
its signals in processes other than the ones expected at hadron
colliders. Similarly, one cannot constrain models in which a singlet
higgs boson $h^0$ (of any mass) couples only to vectorlike leptons.

\begin{center}
{\bf  ~ Acknowledgements}
\end{center}
\normalsize\rm

The authors are grateful to Debajyoti Chowdhury for pointing out an
important error in a previous draft. They have also profited from
discussions with Gautam Bhattacharyya. The authors acknowledge partial
financial support from the University Grants Commission, India (AD) as
well as the Council for Scientific and Industrial Research and
Department of Science and Technology, Government of India (AR). The
work of SR is supported by a project (DO NO. SR/SY/P-08/92) of the
Department of Science and Technology, Government of India. SC would
like to thank the Distributed Information Centre (DIC), Bose Institute,
Calcutta, for computational facilities.

\newpage
\begin{center}
{\bf\large Table 1} $$
\begin{array}{|c|c|c|c|c|c|}
\hline
{\rm Scenario} & ~f~ & \omega_{\gamma f} & \omega_{Zf} & \omega_{gf} &
\omega_{Wf}  \\
\hline
1 & {\rm quark~doublet} & -Q_f & -Q_f\tan\theta_W + 2T_{3f}{\rm
csc}2\theta_W & g_s/e & {\rm csc}\theta_W/\sqrt{2}  \\
\hline
2,3 & {\rm singlet~quark} & -Q_f & -Q_f\tan\theta_W & g_s/e & 0  \\
\hline
4 & {\rm lepton~doublet} & -Q_f & -Q_f\tan\theta_W + 2T_{3f}{\rm
csc}2\theta_W & 0 & {\rm csc}\theta_W/\sqrt{2}  \\
\hline
5 & {\rm singlet~lepton} & -Q_f & -Q_f\tan\theta_W & 0 & 0  \\
\hline
\end{array}
$$
\end{center}
\vskip 20pt
\begin{center}
{\bf\large Table 2} $$
\begin{array}{|c|c|c|c|c|c|c|c|}
\hline
{\rm Scenario} & ~f~ & C_{ \gamma \gamma} & C_ {\gamma Z^{0}} & C_ {Z^0
Z^0} & C_{g_{a} g_{b}} & C_{WW} & C_{ff}  \\
\hline
1 & {\rm quark~doublet} & 3 & 3 & 3 & {1 \over 2} \delta _{ab} & 3 &
\sqrt 3 \\
\hline
2,3 & {\rm singlet~quark} & 3 & 3 & 3 & {1 \over 2} \delta _{ab} & 0 &
\sqrt 3
\\
\hline
4 & {\rm lepton~doublet} & 1 & 1 & 1 & 0 & 1 & 1 \\
\hline
5 & {\rm singlet~lepton} & 1 & 1 & 1 & 0 & 0 & 1 \\
\hline
\end{array}
$$
\end{center}
\newpage
\begin{center}
{\bf\large Table 3}
\end{center}
\begin{center}
\begin{tabular}{|c|c|c|c|c|}
\hline
\multicolumn{5}{|c|}{Higgs width in GeV} \\
\hline
& \multicolumn{2}{|c}{$m_f$ = 50 GeV} &
\multicolumn{2}{|c|}{$m_f$ = 150 GeV}  \\
 \cline{2-5} {Scenario} & \multicolumn{1}{|c}{$m_h$ = 100 GeV} &
\multicolumn{1}{|c|}{$m_h$ = 200 GeV} & \multicolumn{1}{|c}{$m_h$ = 100
GeV} & \multicolumn{1}{|c|}{$m_h$ = 200 GeV} \\ \hline
\multicolumn{1}{|c|}{Doublet Quarks}
& \multicolumn{1}{|c}{0.2058} & \multicolumn{1}{|c|}{72.854} &
\multicolumn{1}{|c}{0.0034} & \multicolumn{1}{|c|}{0.0357}   \\ \hline
\multicolumn{1}{|c|}{Singlet U Quark}
& \multicolumn{1}{|c}{0.0515} & \multicolumn{1}{|c|}{36.289} &
\multicolumn{1}{|c}{0.00085} & \multicolumn{1}{|c|}{0.0088}   \\ \hline
\multicolumn{1}{|c|}{Singlet D Quark}
& \multicolumn{1}{|c}{0.0513} & \multicolumn{1}{|c|}{36.288} &
\multicolumn{1}{|c}{0.00084} & \multicolumn{1}{|c|}{0.00874}   \\
\hline 
\multicolumn{1}{|c|}{Doublet Leptons}
& \multicolumn{1}{|c}{8.97 $\times$ $10^{-5}$} &
\multicolumn{1}{|c|}{24.128} & \multicolumn{1}{|c}{2.126 $\times$
$10^{-6}$} & \multicolumn{1}{|c|}{9.39 $\times$ $10^{-5}$}   \\ \hline
\multicolumn{1}{|c|}{Singlet Lepton}
& \multicolumn{1}{|c}{8.90 $\times$ $10^{-5}$} &
\multicolumn{1}{|c|}{12.063} & \multicolumn{1}{|c}{2.124 $\times$
$10^{-6}$} & \multicolumn{1}{|c|}{2.91 $\times$ $10^{-5}$}   \\ \hline
\multicolumn{1}{|c|}{`Standard' Model} & \multicolumn{1}{|c}{0.0034} &
\multicolumn{1}{|c|}{2.897} &
\multicolumn{1}{|c}{0.0028} & \multicolumn{1}{|c|}{1.511}   \\ \hline
\end{tabular}

\end{center}

\newpage
\begin{center}
{\bf {\large ~ Table Captions}}
\end{center}
\normalsize\rm

{\bf Table 1} : Overall factors multiplying the production (and decay)
matrix element for different scenarios. The symbol $f$ stands for
vectorlike fermions generically. $Q_f$ and $T_{3f}$ denote,
respectively, the electric charge and the third component of weak
isospin of the vectorlike fermion.

{\bf Table 2} : Colour factors multiplying the production (and decay)
matrix element for different scenarios. The symbol $f$ stands for
vectorlike fermions generically. $C_{ij}$ denotes the relevant colour
factors for the decay $h^0 \rightarrow V_iV_j$ through a vectorlike
fermion loop.

{\bf Table 3} : The width of a singlet higgs for $m_h =$ 100 and 200
GeV in various vectorlike fermion scenarios with $\xi = \eta = 1$. For
comparison, the width of a `Standard' model higgs with a hypothetical
top quark of mass 50 and 150 GeV are also shown.

\newpage
\begin{center}
{\bf {\large  ~ Figure Captions}}
\end{center}
\normalsize\rm

{\bf Fig. 1:}   Feynman diagram for (a) the hadroproduction of an $h^0$
and (b) the loop-induced decay of $h^0$ to a pair of vector bosons.

{\bf Fig. 2:}   $h^0$ production rate for a doublet of vectorlike
quarks with (a) $\eta$ = 1, $\xi$ = 1, (b) $\eta$ = 0, $\xi$ = 1, and
(c) $\eta$ = 1, $\xi$ = 0.  The rate for a singlet U type vectorlike
quark with $\eta$ = 1, $\xi$ = 1 is shown in (d) while that for a
doublet of vectorlike leptons with $\eta$ = 1, $\xi$ = 1 is presented
in (e). Upper lines correspond to $m_f = 50$ GeV,  and  lower lines
correspond to $m_f = 200$ GeV; high luminosity options are taken both
for the LHC (solid lines) and the Tevatron* (dashed lines).

{\bf Fig. 3:}   Branching ratios for $h^0$ decay with (a) a doublet of
vectorlike quarks and (b) a singlet vectorlike $U$ quark.  The
conventions followed for each of these is the following: ($i$)   solid:
$h^0
\rightarrow \gamma \gamma$ mode; ($ii$)  solid with dots: $h^0
\rightarrow Z^0 \gamma$ mode; ($iii$) large dashes: $h^0 \rightarrow
Z^0 Z^0$ mode; ($iv$)  dot-dashed:  $h^0 \rightarrow W^+W^-$ mode;
($v$)   small dashes: $h^0 \rightarrow f \bar f$ mode; ($vi$)  dotted:
$h^0 \rightarrow gg$ mode.  We have set $m_f = 100$ GeV throughout.

{\bf Fig. 4:}  $\gamma\gamma$ signal as a function of invariant mass of
the final state in bins of 10 GeV with (a) a doublet of vectorlike
quarks and (b) a singlet vectorlike $U$ quark, for $m_f = 50, 100, 200$
GeV (solid, large dashed and small dashed curves respectively) at the
LHC with the high luminosity option.  The histogram shows the square
root of the number of events from the SM background in each bin.

{\bf Fig. 5:}  $\gamma\gamma$ signal as a function of invariant mass of
the final state in bins of 10 GeV with (a) a doublet of vectorlike
quarks and (b) a singlet vectorlike $U$ quark, for $m_f = 50, 100, 200$
GeV (solid, large dashed and small dashed curves respectively) at the
Tevatron*.  The histogram shows the square root of the number of events
from the SM background in each bin.

{\bf Fig. 6:}  $\ell^+ \ell^- \gamma$ signal as a function of invariant
mass of the final state in bins of 10 GeV with (a) a doublet of
vectorlike quarks and (b) a singlet vectorlike $U$ quark, for $m_f =
50, 100, 200$ GeV (solid, large dashed and small dashed curves
respectively) at the LHC.  The histogram shows the square root of the
number of events from the SM background in each bin.

{\bf Fig. 7:} $\ell^+ \ell^- \ell^+ \ell^-$ signal as a function of
invariant mass of the final state in bins of 10 GeV with a doublet of
vectorlike quarks for $m_f = 50, 100, 200$ GeV (solid, large dashed and
small dashed curves respectively) at the LHC with the high luminosity
option.  The histogram shows the square root of the number of events
from the SM background in each bin.

\newpage
\thicklines
\thispagestyle{empty}
\setlength{\unitlength}{1pt}
\begin{picture}(400,400)(0,0)
\setlength{\unitlength}{1pt}
\put(0,250){\line(1,0){100}}
\put(0,248){\line(1,0){100}}
\put(0,150){\line(1,0){100}}
\put(0,152){\line(1,0){100}}
\put(100,248){\line(4,1){50}}
\put(100,248){\line(3,1){50}}
\put(100,248){\line(2,1){50}}
\put(100,148){\line(4,-1){50}}
\put(100,148){\line(3,-1){50}}
\put(100,148){\line(2,-1){50}}
\multiput(105,250)(10,0){10}{\oval(5,5)[t]}
\multiput(100,250)(10,0){10}{\oval(5,5)[b]}
\multiput(105,150)(10,0){10}{\oval(5,5)[t]}
\multiput(100,150)(10,0){10}{\oval(5,5)[b]}
\multiput(200,-50)(10,0){10}{\oval(5,5)[t]}
\multiput(205,-50)(10,0){10}{\oval(5,5)[b]}

\multiput(200,-150)(10,0){10}{\oval(5,5)[t]}
\multiput(205,-150)(10,0){10}{\oval(5,5)[b]}
\multiput(250,200)(10,0){11}{\line(1,0){5}}
\multiput(50,-100)(10,0){11}{\line(1,0){5}}
\put(200,150){\line(0,1){100}}
\put(200,150){\line(1,1){50}}
\put(200,250){\line(1,-1){50}}
\put(200,-150){\line(0,1){100}}
\put(150,-100){\line(1,-1){50}}
\put(150,-100){\line(1,1){50}}
\put(200,150){\vector(0,1){50}}
\put(200,250){\vector(1,-1){25}}
\put(250,200){\vector(-1,-1){25}}
\put(200,-50){\vector(0,-1){50}}
\put(150,-100){\vector(1,1){25}}
\put(200,-150){\vector(-1,1){25}}
\put(100,149){\circle*{10}}
\put(100,249){\circle*{10}}
\put(200,149){\circle*{7}}
\put(200,249){\circle*{7}}
\put(250,199){\circle*{7}}
\put(200,-149){\circle*{7}}
\put(200,-49){\circle*{7}}
\put(150,-99){\circle*{7}}
\put(50,260){\LARGE $p$}
\put(50,138){\LARGE $p$}
\put(300,205){\LARGE $h^0$}
\put(100,-95){\LARGE $h^0$}
\put(130,230){\LARGE $V_i$}
\put(130,160){\LARGE $V_j$}
\put(235,-175){\LARGE $V_i^\prime$}
\put(235,-35){\LARGE $V_j^\prime$}
\put(180,200){\Large $f$}
\put(230,230){\Large $f$}
\put(230,165){\Large $f$}
\put(210,-100){\Large $f$}
\put(155,-80){\Large $f$}
\put(155,-130){\Large $f$}
\put(160,95){\LARGE Fig. 1a}
\put(180,-205){\LARGE Fig. 1b}
\end{picture}
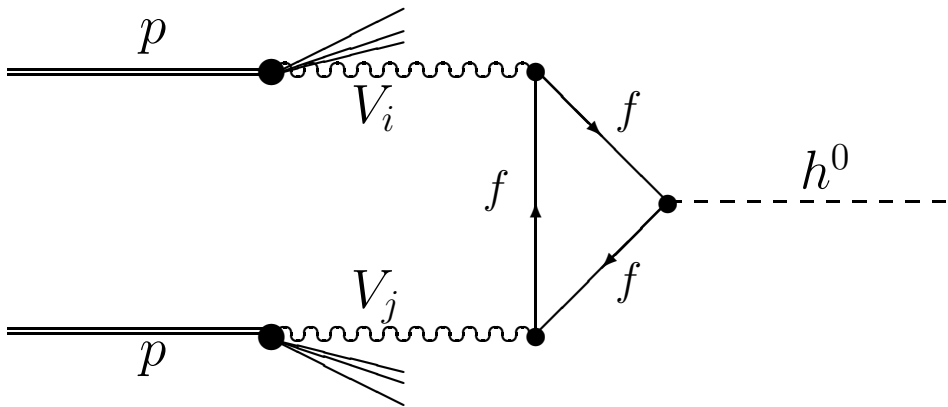
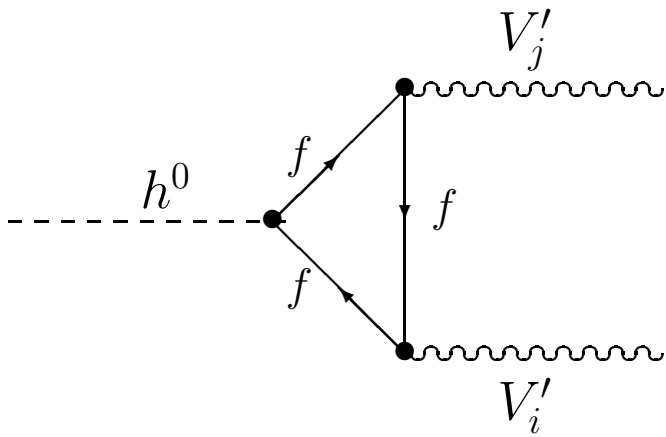
\end{document}